\documentclass[aps,prb,twocolumn,showpacs,preprintnumbers,
superscriptaddress,
amsmath,amssymb,longbibliography]{revtex4-2}

\usepackage[usenames,dvipsnames,svgnames,table]{xcolor}

\definecolor{darkgreen}{rgb}{0,0.5,0}

\newcommand{\ket}[1]{|#1\rangle}

\newcommand{\be}{\begin{equation}}
\newcommand{\ee}{\end{equation}}
\newcommand{\bea}{\begin{eqnarray}}
\newcommand{\eea}{\end{eqnarray}}

\newcommand{\e}{\varepsilon}
\newcommand{\w}{\omega}
\newcommand{\s}{\sigma}
\newcommand{\G}{\Gamma}
\newcommand{\up}{\uparrow}
\newcommand{\down}{\downarrow}

\newcommand{\dk}{d^\dagger}

\newcommand{\ck}{c^\dagger}

\newcommand{\new}[1]{{\color{black}{#1}}}

\newcommand{\exch}{\Delta\e_{\rm exch}}
\newcommand{\exchp}{\Delta\e_{\rm exch}^p}

\usepackage[dvipsnames]{xcolor}

  \newcommand{\Sec}[1]{Sec.~\ref{#1}}
  \newcommand{\App}[1]{App.~\ref{#1}}
  \newcommand{\Eq}[1]{Eq.~\eqref{#1}}

  \newcommand{\Fig}[1]{Fig.~\ref{#1}}

\newcommand{\aw}[1]{{\color[rgb]{.0,.5,.2}{#1}}}

\def\iu{\ensuremath{\mathrm{i}\mkern1mu}}

\usepackage{ulem}
\usepackage{verbatim} 
\usepackage{amsmath}
\usepackage{graphicx}
\usepackage{dcolumn}
\usepackage{bm}

\definecolor{oucrimsonred}{rgb}{0.6, 0.0, 0.0}

\usepackage[colorlinks, citecolor={blue!50!black}, urlcolor={blue!50!black}, linkcolor={red!50!black}]{hyperref}

\begin{document}



\title{Nonequilibrium spintronic transport through Kondo impurities}

\author{Anand Manaparambil}
\email{anaman@amu.edu.pl}
\affiliation{Institute of Spintronics and Quantum Information, Faculty of Physics, 
Adam Mickiewicz University, Uniwersytetu Pozna\'nskiego 2, 61-614 Pozna\'n, Poland}

\author{Andreas Weichselbaum}
\affiliation{Department of Condensed Matter Physics and Materials Science,
Brookhaven National Laboratory, Upton, New York 11973-5000, USA}

\author{Jan von Delft}
\affiliation{Arnold Sommerfeld Center for Theoretical Physics,
    Center for NanoScience, and Munich Center for Quantum Science and Technology,
    Ludwig-Maximilians-Universit\"at M\"unchen, 80333 Munich, Germany}

\author{Ireneusz Weymann}
 \affiliation{Institute of Spintronics and Quantum Information, Faculty of Physics, 
Adam Mickiewicz University, Uniwersytetu Pozna\'nskiego 2, 61-614 Pozna\'n, Poland}

\date{\today}


\begin{abstract}
In this work we analyze the nonequilibrium transport through a quantum impurity (quantum dot or molecule)
attached to ferromagnetic leads by using a hybrid numerical renormalization group-time-dependent density matrix renormalization group thermofield quench approach.For this, we study the bias dependence of the differential conductance through the system, which shows a finite zero-bias peak, characteristic of the Kondo resonance and reminiscent of the equilibrium local density of states.
In the non-equilibrium settings,  the resonance in the differential conductance is also found
to decrease with increasing the lead spin polarization.
The latter induces an effective
exchange field that lifts the spin degeneracy of the dot level.
Therefore as we 
demonstrate,  the Kondo resonance can be restored
by counteracting the exchange field with a finite
external magnetic field applied to the system.
Finally, we investigate the influence of temperature on the nonequilibrium conductance, focusing on the split Kondo resonance.
Our work thus provides an accurate quantitative description of the spin-resolved transport properties relevant for quantum dots and molecules embedded in magnetic tunnel junctions.
\end{abstract}

\maketitle

\section{\label{sec:level1}Introduction}

Charge and spin transport through nanostructures such as nanowires,
quantum dots or  molecules have been under rigorous experimental
as well as theoretical research worldwide.
These studies are motivated primarily by the possible applications
in spintronics, nanoelectronics and spin caloritronics,
as well as fascinating physics emerging at the nanoscale
\cite{Zutic2004Apr,Bauer2012May,Awschalom2013Mar,Hirohata2020Sep}.
In particular, the high research interest in transport through 
artificial quantum impurity systems stems
from the observation of the Kondo effect, a many-body phenomenon,
in which the spin of a quantum impurity
becomes screened by conduction electrons of attached electrodes
\cite{Hewson_1993,Goldhaber-Gordon1998Jan, Cronenwett1998Jul}.
Many studies, both experimental and theoretical ones, focused
on providing a deep understanding of the interplay between the Kondo physics
and other many-body phenomena,
such as e.g. ferromagnetism \cite{Martinek2003Sep,Pasupathy2004Oct} or superconductivity \cite{Yazdani1997Mar,Franke2011May}, have been carried out.
In this regard, especially interesting in the context of spin nanoelectronics,
are quantum dots or molecules attached to ferromagnetic electrodes \cite{Seneor2007Apr,barnasJPCM08}.
\new{Besides the fact that such nanostructures allow for implementing 
devices with high spin-resolved properties}, they enable 
the exploration of the interplay between the itinerant ferromagnetism
with the strong electron correlations \cite{Pasupathy2004Oct,Hamaya2007Dec,Hauptmann2008May,Gaass2011Oct}.
In fact, the spintronic transport properties
of ferromagnetic quantum impurity systems have been a subject of extensive investigations
\cite{Martinek2003Sep,Martinek2003Dec,Lopez2003Mar,Pasupathy2004Oct,Utsumi2005Jun,Swirkowicz2006May,Hamaya2007Dec,Hauptmann2008May,Weymann2010Mar,Gaass2011Oct,Zitko2012Apr,Wojcik2015Apr,Weymann2018Feb,Bordoloi2020Aug},
however, their accurate quantitative description in truly nonequilibrium settings
still poses a formidable challenge.

Reliable equilibrium and linear-response studies of transport through quantum impurity systems
have been made possible by a robust non-perturbative
numerical renormalization group (NRG) method \cite{Wilson1975Oct, Bulla2008Apr}.
Unfortunately, this method falls short when describing the nonequilibrium behavior.
On the other hand, although nonequilibrium situations can be studied
by various analytical methods, their main drawback is an approximate
treatment of electron correlations. It is important to note that
these disadvantages have been overcome by
the time-dependent density matrix renormalization group (tDMRG) method \cite{Schollwock2011}
which however has the drawback that it can only reliably study
the system's behavior for timescales of the order $1/D$,
where $D$ is the half-bandwidth of the conduction band.
A reliable quantum quench approach to study the transport
through quantum impurity systems out-of-equilibrium has been recently proposed by F. Schwarz et al. [\onlinecite{Schwarz2018Sep}].
This approach combines both the NRG and tDMRG methods and, in addition, makes use of the thermofield treatment \cite{deVega2015Nov} to efficiently describe the system.

In this paper, by employing the hybrid NRG-tDMRG thermofield quench approach \cite{Schwarz2018Sep}, we provide an accurate theoretical investigation of the nonequilibrium transport through a quantum impurity interacting with ferromagnetic leads.
In particular, we study the bias voltage dependence of the differential conductance, which exhibits a zero-bias peak, a characteristic feature of the Kondo effect,
when the system is tuned to the particle-hole symmetry point.
We show that the Kondo energy scale in the applied bias potential
decreases with increasing the lead spin polarization.
On the other hand, when we detune the system away from
this symmetry point, we observe a splitting of the zero-bias peak   
for finite lead spin polarization, which can be attributed to the emergence of a local exchange field in the impurity.
Furthermore, we study the behavior of this split-Kondo peak
under external parameters, such as applied magnetic field or temperature. We show that a particular value of \new{the} magnetic field can lead to the restoration of the Kondo resonance in the system.
Moreover, we determine the temperature dependence of the differential conductance at the bias voltage corresponding to the split Kondo peak. 

The paper is organized as follows. In \Sec{Sec:Model} we describe the model and method used in calculations. Main results and their discussion are presented in \Sec{Sec:Results}, where we first analyze the differential conductance at the particle-hole symmetry point, and then study the effect of finite exchange field on the transport behavior. We also examine the possibility to restore the Kondo effect by magnetic field and determine the temperature dependence. Finally, the paper is summarized in \Sec{Sec:Summary}. 

\section{Model and method}
\label{Sec:Model}

\begin{figure}
    \centering
\includegraphics[width=0.9\columnwidth]{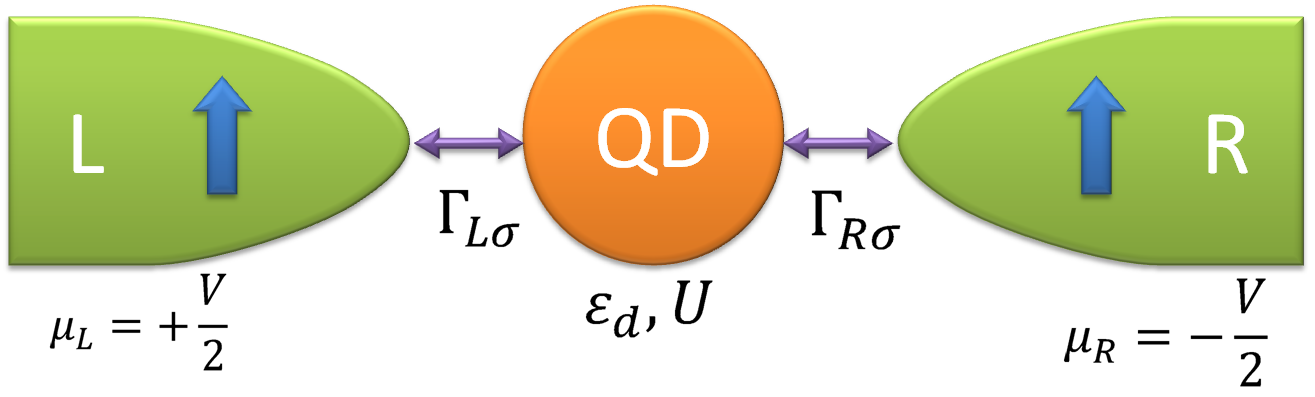}
\caption{
Model system -- A magnetic impurity (quantum dot or molecule), characterized by 
an orbital level of energy $\e_d$ and Coulomb correlations $U$,
is attached to two ferromagnetic contacts with spin-dependent
coupling strengths $\Gamma_{L\sigma}$ and $\Gamma_{R\sigma}$, respectively.
These leads are locally in equilibrium at a global temperature $T$,
yet with a voltage bias $V=\mu_L - \mu_R$ that 
is applied symmetrically across them.
}
\label{Fig:schematic}
\end{figure}

The considered system consists of a quantum impurity (quantum dot or a molecule)
attached to two ferromagnetic leads with spin-dependent couplings,
subject to a voltage bias, as shown schematically in Fig.~\ref{Fig:schematic}.
More specifically, such system can be described by a single impurity Anderson model \cite{anderson_LocalizedMagnetic_1961}, in which the quantum impurity is modeled as
\be
H_{\text{imp}} 
 = \sum_\sigma \e_{d\sigma} n_{\sigma} 
  + U{n}_{\up}{n}_{\down},
\label{eq:Himp}
\ee
with ${n}_\s = \dk_\s d_\s$,
where $\dk_\s$ creates an electron
with spin $\sigma \in \{\up,\down\}
\equiv \{+1,-1\}$ at 
the impurity, 
$\e_{d\sigma} \equiv \e_d  - \tfrac{\sigma}{2} B$
denotes the energy of an impurity energy level
with $B$ the external magnetic field 
in units of $g\mu_B\equiv 1$, and
$U$ the Coulomb repulsion experienced when the level is doubly occupied.

 The leads attached to the impurity are assumed to be ferromagnetic metals
 and are characterized by the Fermi functions, $f_\alpha (\w) = [e^{(\w-\mu_\alpha)/T} +1]^{-1}$ (using units 
 $\hbar=k_B=e=1$, throughout),
 where the index $\alpha$ refers to the leads,
 $\alpha \in \{L,R\} \equiv \{-1,+1\}$
 and $\mu_\alpha = \alpha V/2$. 
 The lead Hamiltonian reads as follows
 \be
 H_{\text{lead}} = \sum_{\alpha k \s}
\e_{\alpha k \s}
\ck_{\alpha k \s } c_{\alpha k \s }^{\ },
 \ee
 with $\ck_{\alpha k \s}$ creating an electron in lead $\alpha$ with energy
%
$\e_{\alpha  k \sigma}$,
 %
%
%
momentum $k$, and spin $\s$.
 The quantum impurity is coupled to the leads
 according to the Hamiltonian $H_{\rm hyb}$,
 \be
 H_{\text{hyb}} = \sum_{\alpha k \s} (v_{\alpha k \s} \dk_\s c_{\alpha k \s} + \text{H.c.}).
 \ee
  Electronic transition between each lead mode $c_{\alpha \s k}$ and the impurity spin state $\s$ is specified by the tunnel matrix elements $v_{\alpha k \s }$.
  This coupling between the lead and impurity induces an impurity-lead
  hybridization in the system, expressed by the hybridization function $\G_{\alpha \s}(\w) = \pi \sum_{k} |v_{\alpha k \s}|^2 \delta(\w - \e_{\alpha k \s})$.
    Finally, the total Hamiltonian of the system reads, 
\be
H_{\text{tot}}= H_{\text{imp}} +  H_{\text{lead}} +  H_{\text{hyb}}.
\ee
In this work we assume a constant
hybridization function 
over the entire bandwidth $2D$ (we use $D:=1$
as unit of energy throughout, unless specified otherwise).
The hybridization function can thus be written as
$\Gamma_{\alpha \sigma}(\omega)
  = \Gamma_{\alpha \sigma} \,\vartheta(D-|\omega|)$, 
  with $\vartheta(\aw{\cdot})$ the Heaviside step function
  and constant $\Gamma_{\alpha \sigma} =
  \pi\rho_{\alpha\s} |v_{\alpha\s}|^2$,
  where $\rho_{\alpha\s}$ is
  the spin-dependent density of states of  lead $\alpha$.
Assuming $v_{\alpha\s} \equiv v$ is independent of spin or lead,
  it is then convenient to introduce the spin polarization
  $p_\alpha$ of the ferromagnetic contact $\alpha$,
\begin{eqnarray}
  p_\alpha
  = \tfrac{\rho_{\alpha\up} - \rho_{\alpha\down}}%
    {\rho_{\alpha\up} + \rho_{\alpha\down}}
\text{ .}\label{eq:p:alpha}
\end{eqnarray}
The coupling strength can be then written
as $\Gamma_{\alpha\s} = (1 +\sigma 
p_\alpha)\Gamma_\alpha$,
  with $\Gamma_{\alpha} = (\Gamma_{\alpha\up}+\Gamma_{\alpha\down})/2$.
  The total coupling strength for spin $\s$ is given by, $\Gamma_\s = \Gamma_{L\s}+\Gamma_{R\s}$.
  In the following we assume that the system is left-right symmetric,
  i.e. $\Gamma_L = \Gamma_R \equiv \Gamma/2$ and $p_L = p_R \equiv p$.
Consequently, the computed electrical current
through the impurity is independent of the
sign of the applied bias voltage $V$, and therefore
it suffices to analyze $V\geq0$.

The impurity parameters are fixed to
\begin{eqnarray}
   U=0.012\ ,\quad \Gamma=0.001
\label{eq:param}
\end{eqnarray}
throughout our paper to ensure a well-defined
Kondo regime well isolated from the finite bandwidth,
with the impurity level position $\varepsilon_d$ varied
from particle-hole symmetric ($\varepsilon_d=-U/2$)
to asymmetric ($\varepsilon_d=-U/3$).
  
We use a hybrid NRG-tDMRG thermofield quench method \cite{Schwarz2018Sep} to study the non-equilibrium behavior of the system. This initializes the leads in thermal equilibrium
at their respective chemical potentials,
before they get dynamically coupled
when smoothly turning on the coupling to the impurity.
This method can treat the correlations exactly while 
sustaining the nonequilibrium conditions of a fixed chemical potential difference and fixed temperature in the leads.
We define a transport window (TW) defined by the Fermi functions of the leads $[f_L(\w) \ne f_R(\w)]$. The energies outside the TW are assumed to be in equilibrium and discretized logarithmically according to the logarithmic discretization parameter $\Lambda$ and energies inside the TW are assumed to be out of equilibrium and discretized linearly according to the linear discretization parameter $\delta$. A thermofield treatment is performed on the discrete energy levels which maps the system to a particle-hole representation. Moreover, in this particle-hole picture, the tunnel matrix elements turn out to be functions of the bias voltage $V$, thus containing the information about the non-equilibrium settings.
  The particle and hole modes in the leads are recombined separately, leaving the impurity coupled with one set of effective particle and one set of effective hole modes. Then, NRG is applied to the logarithmically discretized part of the system, resulting in a renormalized impurity (RI), which is coupled to the linearly discretized part of the hole and particle chain. We represent the RI in the matrix product state (MPS) framework as one site of the MPS chain coupled to completely filled particle and completely empty hole modes in the linearly discretized sector. The system is then time-evolved using a second-order Trotter time evolution, where the coupling between the RI and the lead modes are switched on over a finite time window. Further details of the method are presented in App.~\ref{app:Appendix}.

\section{Results and discussion}
\label{Sec:Results}

In the case of quantum dots or molecules attached to ferromagnetic
contacts the transport properties are strongly dependent
on the spin-resolved charge fluctuations between the impurity
and ferromagnets. These fluctuations give rise to the level renormalization
$\delta \e_\sigma$. Because for $p>0$,
$\delta\e_\uparrow \neq \delta\e_\downarrow$,
a spin splitting of the impurity level can be generated,
$\exch \equiv \delta\e_\uparrow - \delta\e_\downarrow$,
referred to as a ferromagnetic-contacted induced exchange field.
Here the exchange field is defined such
that $\exch > 0$ tends towards a {\it negative}
impurity magnetization, which in terms of sign is 
contrary to the definition of $B$ in \Eq{eq:Himp}. 
Hence the effective total magnetic field experienced
by the impurity is given by
\be
    B_\mathrm{tot}^\mathrm{eff} \simeq
    B - \exch
\label{eq:Beff}
\ee
The exchange field in the local moment regime can be estimated within
the second-order perturbation theory and it is given by \cite{Martinek2003Sep}
\be \label{Eq:exch}
\exch (p) = \tfrac{2 p \G}{\pi} \text{Re} \big[
\phi(\e_d) - \phi(\e_d + U)
\big],
\ee
where
$\phi(\varepsilon)
= \Psi(\tfrac{1}{2} + \tfrac{\iu\varepsilon}{2\pi T} )$,
with $\Psi(z)$ being the digamma function.
At $T=0$, the formula for the exchange field simply becomes 
\begin{eqnarray}
   \exchp = p \underbrace{
   \tfrac{2 \G}{\pi} \ln | \tfrac{\e_d}{\e_d+U} |}_{
   \equiv \gamma(\Gamma, \frac{\varepsilon_d}{U})}
\text{ .}\label{Eq:exch:0}
\end{eqnarray}
The most important property of $\exch$
is its tunability with changing the position of the orbital level.
As follows from the above formula, $\exch$ changes
sign when crossing the particle-hole (p-h) symmetry point, $\e_d=-U/2$, at which it vanishes.

We begin our analysis with the study of the influence of the lead polarization
on the nonequilibrium conductance of the system
when the impurity energy level is tuned to $\e_d=-U/2$.
We then proceed to examine the case when the system is detuned
from the p-h symmetry point ($\e_d\neq -U/2$),
where the exchange field can introduce spin-splitting in the system.
We also analyze the influence of temperature and applied magnetic field
on the split Kondo resonance observed in the differential
conductance out of the p-h symmetry point.

\subsection{Conductance at the p-h symmetry point}

\begin{figure}[t]
    \includegraphics[width=0.95\columnwidth]{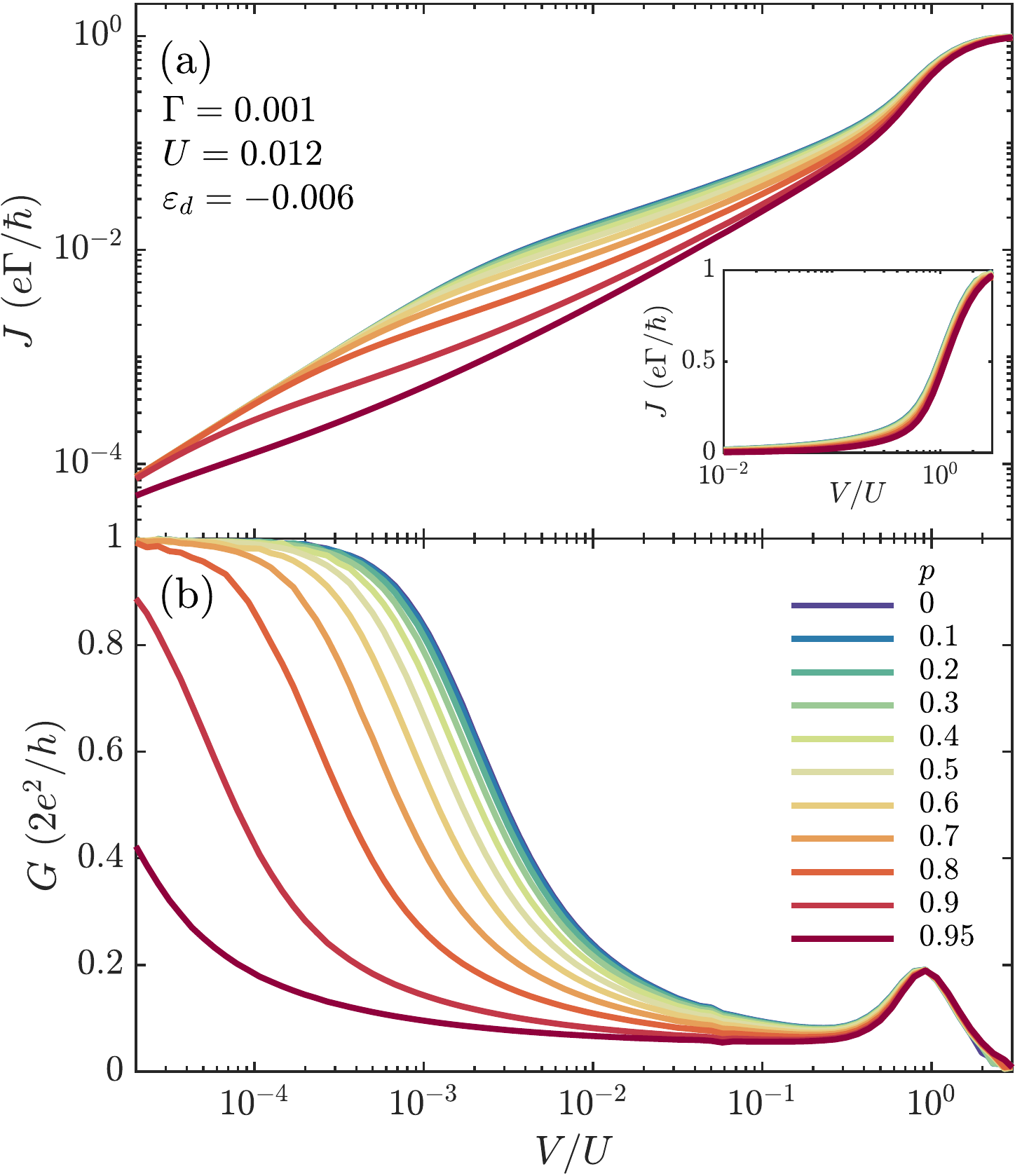}
    \caption{
    The bias voltage dependence
    at $T=0$ and particle-hole symmetry
    $\varepsilon_d = -U/2$ of (a) the mean
    current $J$ on a log-log scale (inset lin-log), and (b) the
    corresponding differential conductance $G$ on a lin-log
    scale.  The various curves are for different values of the
    lead spin polarization $p$, as indicated.
}
\label{Fig:ph}
\end{figure}

\begin{figure}[t]
\includegraphics[width=0.95\columnwidth]{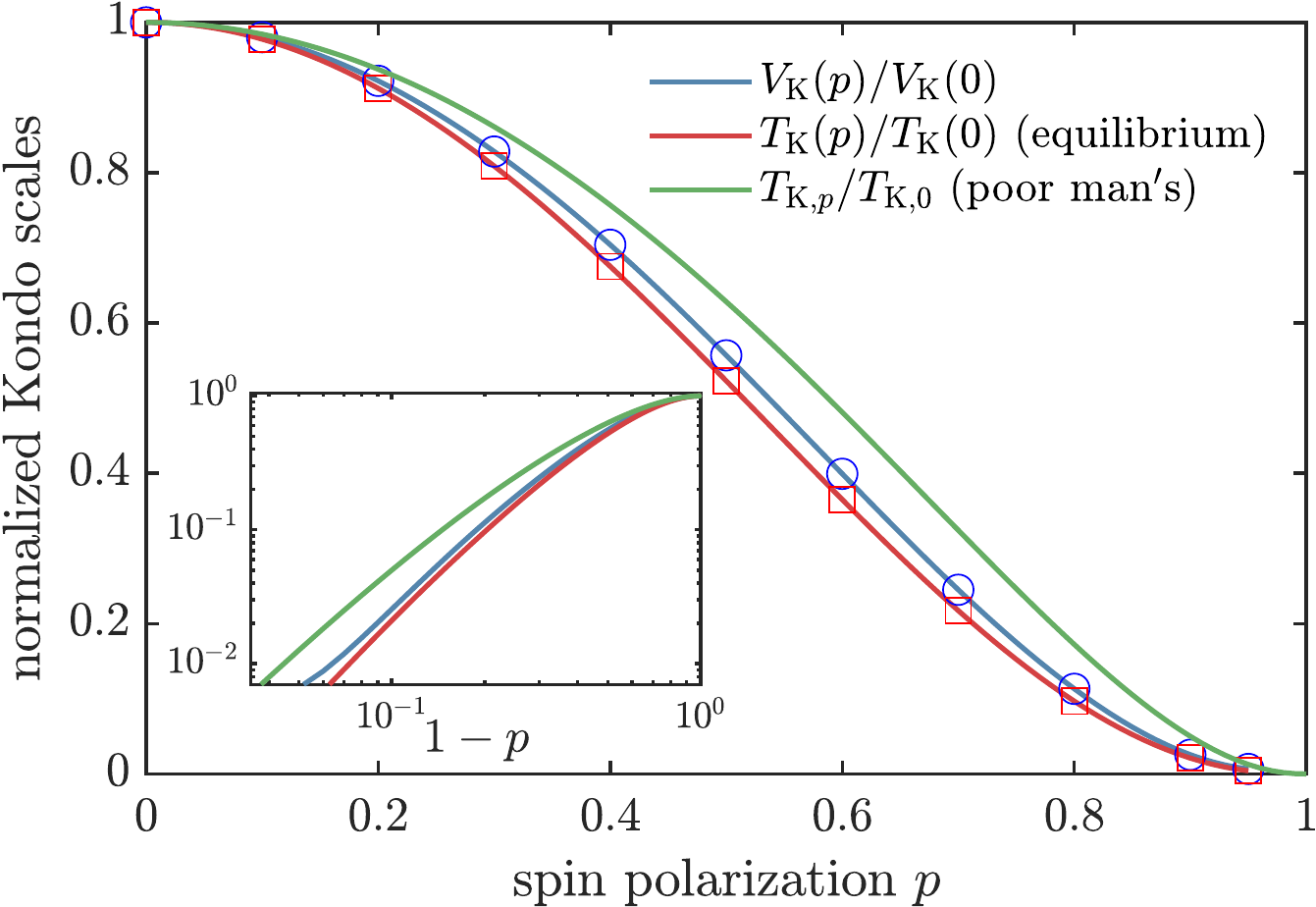}
\caption{
    Comparison of the Kondo energy scale $V_{\mathrm{K}}$
    in the applied bias potential at $\varepsilon_d=-U/2$
    [\Eq{eq:param}] with the corresponding
    equilibrium values of $T_{\mathrm{K}}$ obtained from NRG calculations,
    and the theoretical prediction for $T_{\mathrm{K}}$ (denoted by $ T_{\mathrm{K},p}$)
    using the poor man's scaling approach, Eq.~(\ref{eq:TKP}).  The Kondo
    energy scales are normalized with their corresponding values
    at $p=0$, having $T_{\mathrm{K},0} = 2.2 \cdot 10^{-5}$ and $V_{K}(0) =
    3.6 \cdot 10^{-5}
    =1.64 \, T_{\mathrm{K},0}$, and $T_{\mathrm{K}}(0) = 2.6 \cdot 10^{-5} = 1.18\, T_{\mathrm{K},0}$.
}
\label{Fig:ph_TK}
\end{figure}

The mean  current $J(V)$ and the corresponding differential
conductance $G(V)$ through the system 
calculated at the particle-hole symmetry point ($\e_d=-U/2$)
for different values of the lead spin polarization $p$ are presented in Fig.~\ref{Fig:ph}.
For this we always evaluate the symmetrized current
as discussed in \App{sec:NRG:tDMRG} [cf. \Eq{eq:J:sym}].
For $p=0$, we observe a zero-bias conductance peak,
characteristic of the Kondo effect \cite{Goldhaber-Gordon1998Jan, Cronenwett1998Jul}.
However when $p$ is finite, the Kondo temperature
is found to decrease with increasing the lead spin polarization.
This was predicted to affect the Kondo temperature
of the system at equilibrium by using the
poor man's scaling method as \cite{Martinek2003Sep}
\be
  T_{\mathrm{K},p} \equiv
\sqrt{\tfrac{\G U}{2}} \exp\Big\{\tfrac{\pi\e_d(\e_d+U)}{2\G U} \tfrac{\text{arctanh}(p)}{p}\Big\}.
\label{eq:TKP}
\ee

The decrease of the Kondo energy scale with spin polarization
can be understood by realizing that by construction with \Eq{eq:p:alpha},
increasing polarization reduces the hybridization of the suppressed spin orientation.
As such, this decreases the rate of spin-flip cotunneling processes
responsible for the Kondo effect.

To quantitatively elucidate the influence of $p$ on the Kondo effect,
we define the Kondo energy scale $V_{\mathrm{K}}$ in the applied bias voltage
as the half maxima point of the conductance curve,
i.e., $G(V_{\mathrm{K}})/G(0) = \frac{1}{2}$ at $T=B=0$.
In Fig.~\ref{Fig:ph_TK} we present the dependence of $V_{\mathrm{K}}$
obtained from our  NRG-tDMRG numerical calculations
along with the Kondo temperature $T_{\mathrm{K},p}$ estimated from
Eq.~(\ref{eq:TKP}) by using the poor man's scaling,
and $T_{\mathrm{K}}(p)$ calculated using the equilibrium NRG \cite{FlexibleDMNRG}
from the temperature dependence of the linear conductance
based on the definition $G(T_{\mathrm{K}})/G(0) =  \frac{1}{2}$.
Our nonequilibrium data corroborates the
general tendency to decrease the Kondo energy
scale with increasing the spin polarization $p$.
However, \Fig{Fig:ph_TK} also demonstrates some deviations:
$V_{\mathrm{K}}$ is slightly larger than the equilibrium $T_{\mathrm{K}}$,
but smaller than the Kondo temperature predicted by the analytical formula (\ref{eq:TKP}),
after normalizing the Kondo energy scales with respect to their respective values at $p=0$.

\subsection{Effect of finite exchange field}

\begin{figure}[t]
    \centering
\includegraphics[width=0.95\columnwidth]{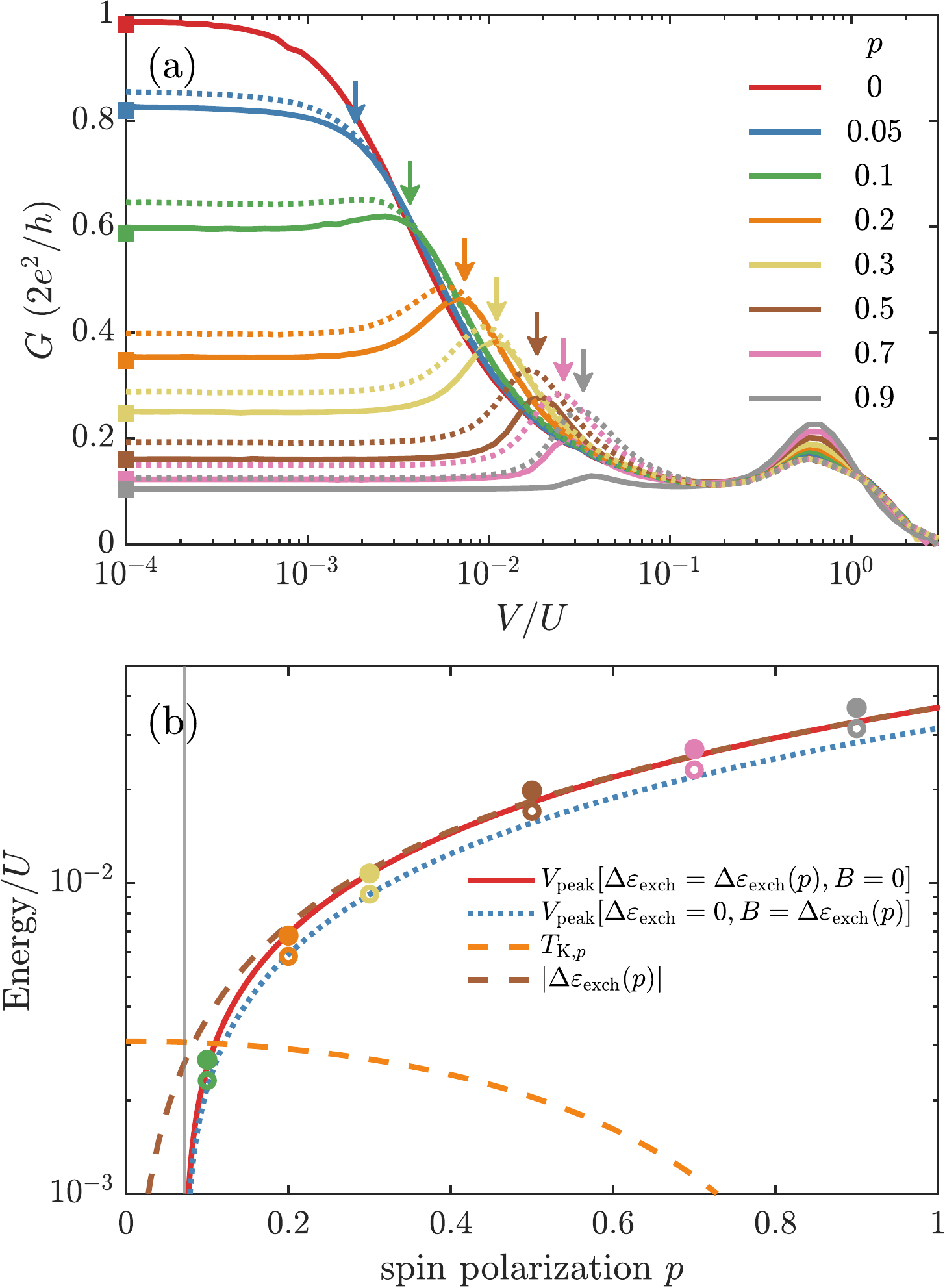}
\caption{ (a)
     The differential conductance $G$ as a function of the bias voltage in the case 
     when the orbital level is detuned from the
     particle-hole symmetry point (solid lines)
     using $\e_d=-U/3$ 
     [\Eq{eq:param}] 
     for different values of the spin polarization $p$ as indicated in
     the legend.
     To check the continuity from the equilibrium regime,
     the corresponding NRG results for the linear response conductance
     are marked by the color-matched symbols (squares) 
     on the left vertical axis.
     For comparison, we also show curves, where the macroscopic
     spin polarization was turned off and replaced,
     instead, by the corresponding local magnetic field $B=\exchp$ (dotted curves).
     Here the value for $\exchp$ was determined
     by \Eq{Eq:exch} at $T=0$,
     and its absolute value is indicated by the color-coded arrows.
     For $p=0$, we obtain $ T_{K0} = 3.7 \cdot 10^{-5}, V_{\mathrm{K}} = 6 \cdot 10^{-5} = 1.62\, T_{\mathrm{K},0}$
     for the voltage bias where the differential conductance drops to half its zero-bias value.
     (b) The filled (empty) circles maps the location of the split-Kondo peak
     from the solid (dotted) curves in Fig.~\ref{Fig:detuned}(a), denoted by  $V_{\mathrm{peak}}$,
     tracked as a function of spin polarization $p$.
     Solid lines in the panel (b) present the extrapolation using a linear fit on the squared data
     for the smallest polarizations (first three data points),thus fitting $V_{\rm peak} = a_0 \gamma \sqrt{p^2 - p_0^2}$
     having $\exch=\gamma p$ with $\gamma=0.4413\,\Gamma$
     [cf. \Eq{Eq:exch:0}],
     and the fit parameters
     $a_0 = 1.001, p_0 = 0.072$ (vertical line).
     Similarly, the fit on the open symbols (dotted line)
     results in $a_0 = 0.860, p_0 = 0.070$. \vspace{-.2in}
     }
\label{Fig:detuned}
\end{figure}

We now discuss the behavior of the differential conductance in
the case when the energy level is away from the p-h symmetry
point ($\e_d=-U/3$), but still in the local moment regime where
the strong electron correlations play a vital role.
The solid lines in Figure~\ref{Fig:detuned}(a) show the bias dependence of the
conductance with increase in the lead spin polarization $p$, 
computed at zero external magnetic field.
One observes a finite zero-bias peak that gets suppressed when
$p$ grows.  This effect can be attributed to the emergence of
exchange field in the system, cf. Eq~(\ref{Eq:exch}).  The
exchange field introduces a spin splitting of the orbital level,
which suppresses the Kondo resonance, once $|\exch|\gtrsim
T_{\mathrm{K}},V_{\mathrm{K}}$.  The color-coded arrows in
Fig.~\ref{Fig:detuned}(a) indicate the magnitude of the exchange field
for the corresponding spin polarizations obtained from
Eq.~(\ref{Eq:exch}) with $T=0$. When the exchange field energy
approaches the Kondo energy scale of the system,
$|\exch|\approx T_{\mathrm{K}}$,
the zero-bias conductance becomes suppressed. 
When increasing the spin polarization further,
the differential conductance starts to develop a peak
around $V\equiv V_{\mathrm{peak}} \approx |\exch|$,
which is a reminiscent of the splitting of the local
density of states (lDOS) vs. frequency in the presence of a
sufficiently strong local magnetic field.
To be specific, the peak in the differential conductance presented
in \Fig{Fig:detuned}(a) emerges  for $p\gtrsim 0.1$.
For this value of spin polarization, one can find that  
$ T_{\mathrm{K},p}= 3.66 \cdot 10^{-5},|\exchp| \simeq 4.4 \cdot 10^{-5} = 1.2 \, T_{\mathrm{K},p}$.
Increasing the polarization further,
the peak at $V \approx |\exch|$ persists while
at the same time, the conductance overall also diminishes.

The dotted lines in Fig.~\ref{Fig:detuned}(a) correspond to the
case in which the system has no exchange field (i.e., $p=0$),
but there is an external magnetic field applied, whose magnitude
equals the exchange field calculated from
Eq.~(\ref{Eq:exch}) according to the spin polarizations
mentioned in Fig.~\ref{Fig:detuned}(a). This comparison shows two
major differences between the exchange field and the magnetic
field. Firstly, a strong enough exchange field 
suppresses the split-Kondo peak in the differential conductance
significantly more strongly and only leaves a residual conductance derived from the
hybridization side peaks energies $V\simeq \varepsilon_d$
[note the log-scale in \Fig{Fig:detuned} (a)]. This is
mainly attributed to the fact that the Kondo scale gets
reduced with increasing the spin polarization [cf. \Fig{Fig:ph_TK}],
such that the ratio $|\exch|/T_{\mathrm{K}}$ is enhanced for the presence
of an exchange field when compared to a local magnetic field. 
Secondly, the location of the split Kondo peak for finite $p$ occurs at
slightly higher voltages than for the case of a local magnetic field. 
The latter effect may be attributed to
$B \approx \exch$ representing a lowest-order estimate.
The explicit dependence of $V_{\mathrm{peak}}$ on the spin polarization $p$
in the two above-discussed cases is shown in Fig.~\ref{Fig:detuned}(b).
For comparison, we also present the $p$-dependence of 
$\exch$ and $T_{\mathrm{K},p}$ estimated from the respective analytical formulas.
One can see that indeed the split Kondo peak emerges when $|\exch|\gtrsim T_{\mathrm{K},p}$.
Moreover, by comparing $\exch$ and $T_{\mathrm{K},p}$,
one can find that these two energy scales become equal for $p=0.0834$.
Keeping in mind that this is an approximate estimate,
our numerical results corroborate this tendency very well. 
The split-Kondo peak shows a slightly nonlinear behavior around low spin polarizations. We fit the $V_{\mathrm{peak}}^2$ data against $p^2$ to unveil any behavior of the form $V_{\mathrm{peak}}
\aw{\sim} 
\sqrt{p^2-p_0^2}$. Both the fits for the exchange field and the corresponding magnetic field 
give essentially the same 
value of $p_0\aw{\simeq 0.071}$ 
indicated by the grey vertical line on panel \Fig{Fig:detuned}(b).
The prefactor of the fit is exactly one ($1.001$) within
numerical accuracy in the presence of polarization,
having $V_{\rm peak} \simeq \sqrt{(\exch)^2 - (\gamma p_0)^2}$
[cf. \Eq{Eq:exch:0}]. 
This is also clearly seen in \Fig{Fig:detuned}(b)
in that the fit exactly coincides with $\exch$
for larger $p$. In the case
of a substitute local magnetic field $B=\exchp$
but unpolarized leads, the fit reads
$V_{\rm peak} \simeq 0.860 \sqrt{(\exch)^2 - (\gamma p_0)^2}$.
This systematically offsets the peaks with the dashed data
in \Fig{Fig:detuned}(a) by a constant factor $0.860$
towards slightly smaller values of the bias voltage,
yet leads to a disappearance of the split-peak at around
the same polarization $p_0$.
On the semilog scale in \Fig{Fig:detuned}(b) this change
in the prefactor simply shifts the fits vertically relative
to each other as also reflected in the data for 
the full polarization range.

The symbols on the left vertical axis in \Fig{Fig:detuned}(a)
correspond to the linear response data obtained by NRG,
which is equivalent to the differential conductance for $V\to0$.
As also seen in later figures, while we have good overall
consistency [e.g., see inset of \Fig{Fig:temperature}(b)],
there are minor quantitative differences in the NRG-tDMRG
results while comparing with the linear-response NRG results.
These are attributed to the different parametrization and 
discretization schemes. Specifically, linear conductance within linear
response in NRG can be obtained strictly at $V=0^+$ \cite{Meir1991Jun}.
In constrast, the NRG-tDMRG approach always must assume
a small but finite voltage in the presence of a finite
level spacing with the objective to numerically compute
a steady-state current via a real-time simulation.

\subsection{The influence of magnetic field}

\begin{figure}[t]
    \centering
\includegraphics[width=0.95\columnwidth]{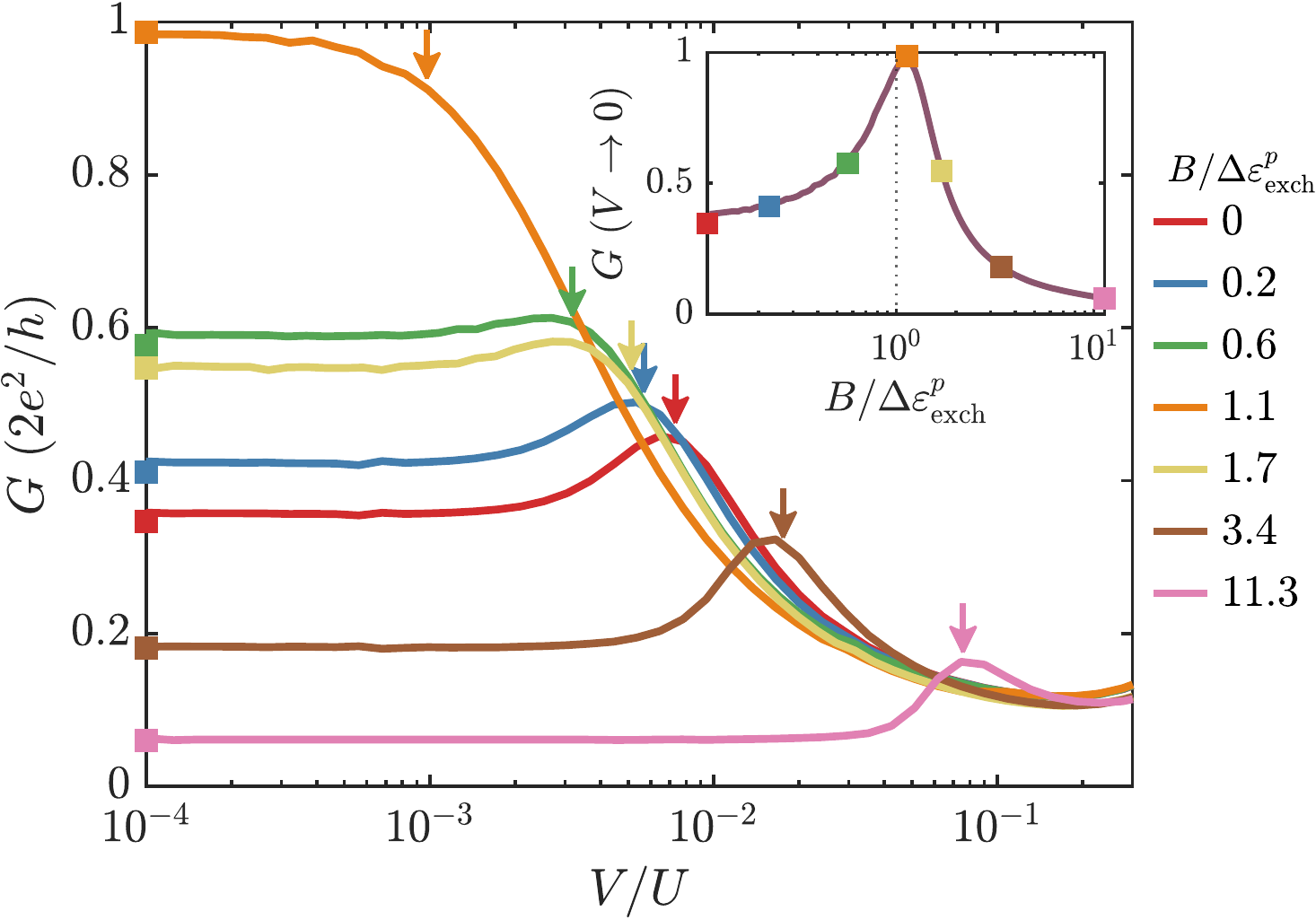}
\caption{
     The differential conductance $G$ as a function of the bias voltage
     calculated at fixed $p=0.2$
     for different values of external magnetic field as indicated
    using $\e_d=-U/3$ [\Eq{eq:param}],
    thus having $\exch^{p} 
     = -0.0882 \, \Gamma$ [\Eq{Eq:exch:0}].
     The color-matched arrows indicate
     $|B_\mathrm{tot}^\mathrm{eff}|$
    as defined in \Eq{eq:Beff}.
     The corresponding NRG results for the linear response conductance are 
    shown by the color-matched symbols (squares)
     on the left vertical axis. The inset shows the behaviour of $G(\aw{B,}V \rightarrow 0)$ as a function of the magnetic field $B$,
     with a significantly more dense set of data points
     from NRG-tDMRG (line), and the symbols from NRG
     as in the main panel.
     The maximum of $G(B)$ occurs
     at $B_{\rm max} 
     = 1.12 \,\exch^{p}$. 
}
\label{Fig:magnetic}
\end{figure}

In Fig.~\ref{Fig:magnetic} we study the influence of external magnetic field
on the split Kondo peak exhibited by the system detuned out of the p-h symmetry point
assuming the lead spin polarization $p=0.2$.
We observe a full restoration of the zero-bias Kondo resonance
by an applied magnetic field with magnitude that can counterbalance
the spin splitting induced by the exchange field, see the curve for \new{$B = 1.1 \, \exchp$} in
Fig.~\ref{Fig:magnetic}. However, a further increase in magnetic field is shown to suppress the zero-bias peak again.
\new{This behavior qualitatively matches the experimental results discussed in the Fig.~2 of the Ref. \cite{Hamaya2007Dec}.}
As seen from the color-coded arrows in \Fig{Fig:magnetic},
the position of the split Kondo resonance
corresponds to $V\approx |B_\mathrm{tot}^\mathrm{eff}|$
as defined in \Eq{eq:Beff}. The revival of the Kondo resonance can be distinctly observed from the inset of \Fig{Fig:magnetic} where $G(V\rightarrow 0)$ exhibits a maximum 
around $B_{\rm max} \simeq \exch^p$ such that
$B_\mathrm{tot}^\mathrm{eff} \simeq 0$ [\Eq{eq:Beff}].
More precisely, from in the inset of \Fig{Fig:magnetic},
$B_{\rm max} 
= 1.12 \, |\exch^{p=0.2}|$,
with the small difference
primarily attributed to the perturbative nature
of the analytic formula \Eq{Eq:exch}.
The prefactor approximately coincides with
a similar scale factor already encountered
with \Fig{Fig:detuned}(b) where $B = \exch^p$
also underestimated the peak position by an
approximate factor $1/0.860 = 1.16$.

\begin{figure}[t]
    \centering
\includegraphics[width=0.95\columnwidth]{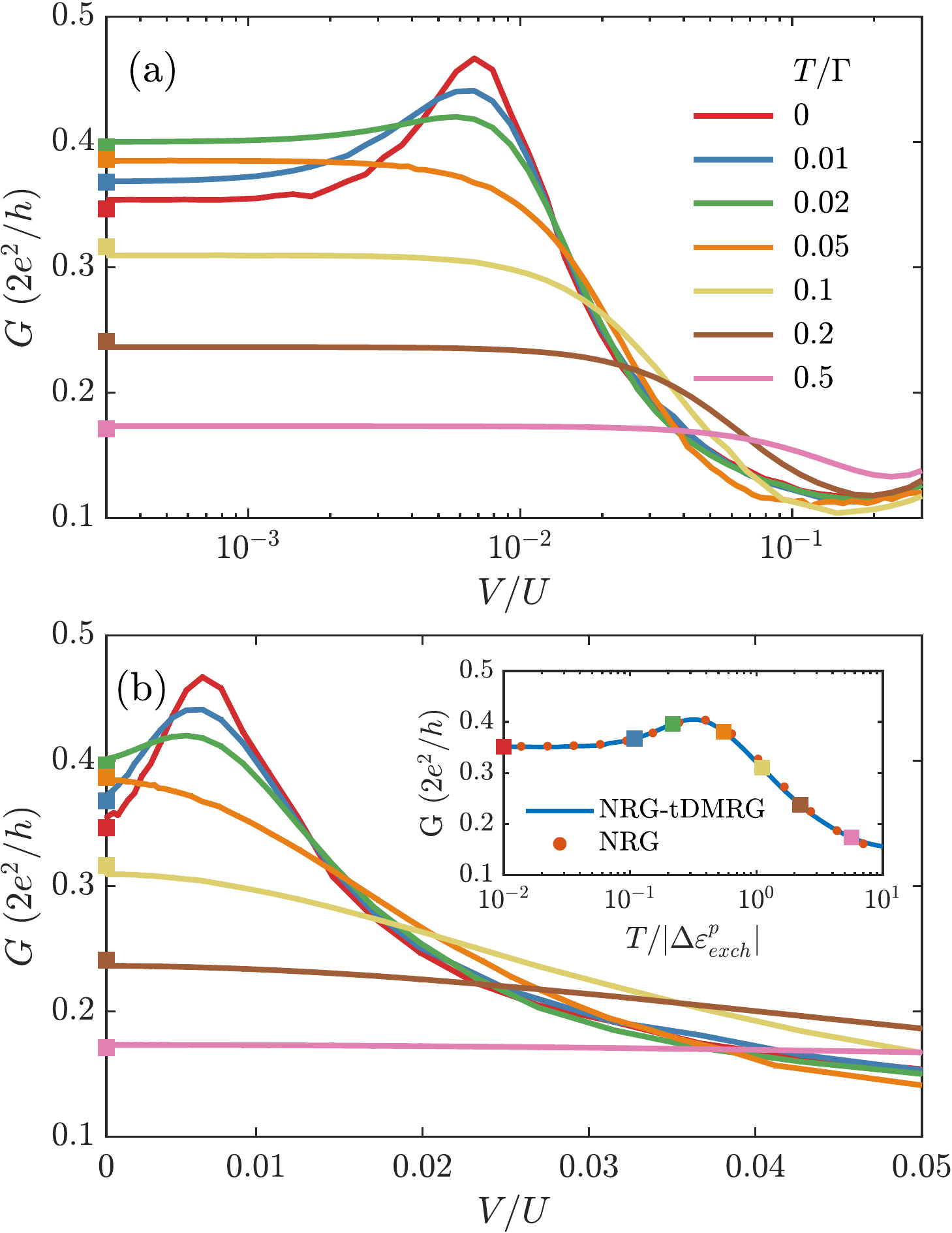}
\caption{
    The bias voltage dependence of the differential conductance $G$
    for fixed $p=0.2$ and $B=0$ calculated for different temperatures     as indicated in the legend,
    using $\e_d=-U/3$ [\Eq{eq:param}].
    Panel (a) presents $G(V)$ on the logarithmic scale,
    while (b) shows the same data
    on the linear scale with focus on the low-bias behavior.
    The linear-response NRG results are shown by the
    color-matched symbols on the left axis.    
    The inset in (b) compares
    the linear response conductance as
    a function of temperature
    of our hybrid scheme (using $V=10^{-3}T$)
    with NRG results, where the large square symbols of the latter
    are identical with the ones in the main panels. 
}
\label{Fig:temperature}
\end{figure}

\begin{figure}[t]
    \centering
\includegraphics[width=0.95\columnwidth]{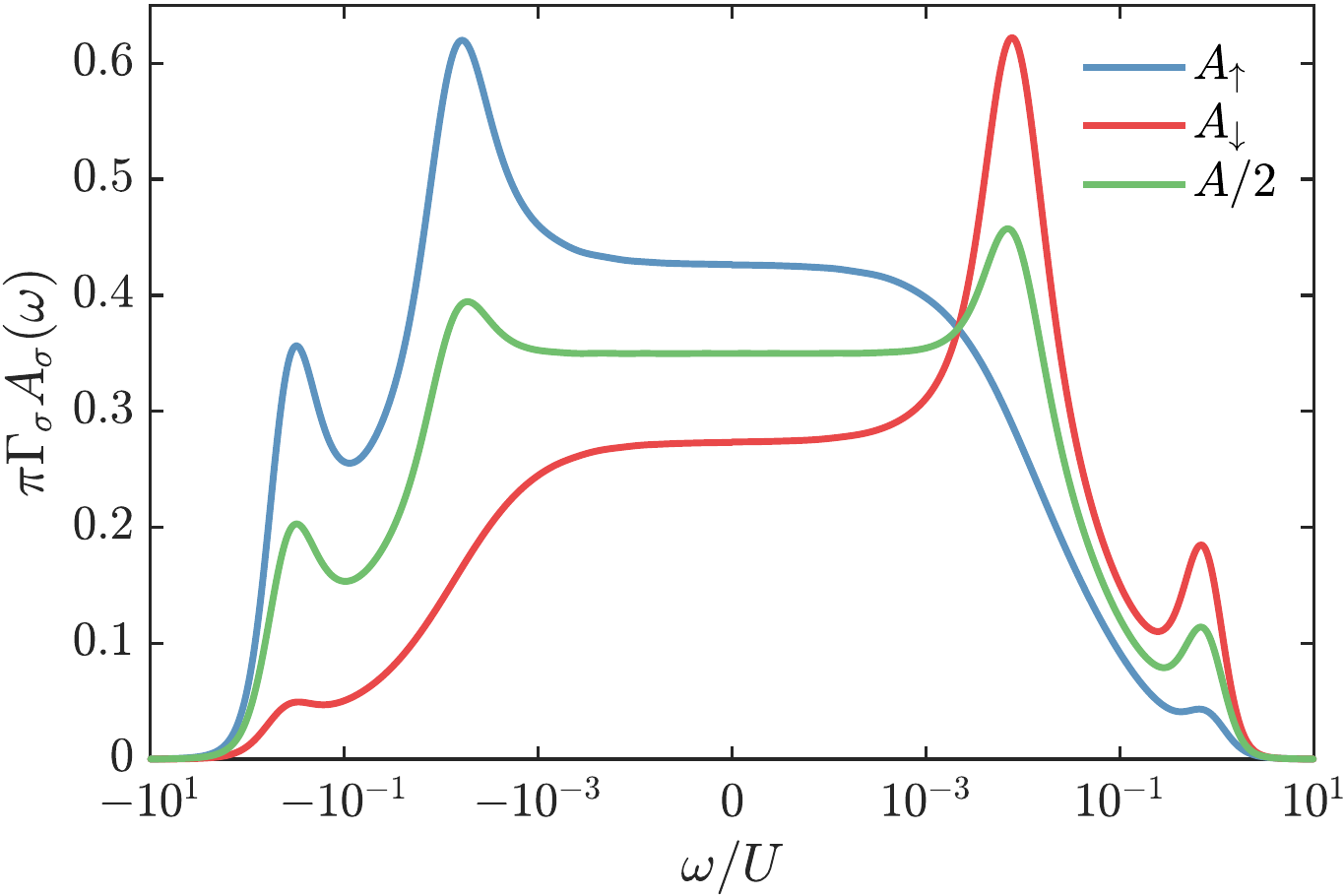}
\caption{
   The energy dependence of the equilibrium zero-temperature
   normalized spectral function $\pi\Gamma_\sigma A_\s(\omega)$
   calculated for $\e_d=-U/3$, $p=0.2$ and $B=0$ [\Eq{eq:param}].
   The dashed lines show the negative
   part of the energy spectrum.
   Note the logarithmic energy scale. }
\label{Fig:Spec}
\end{figure}

\begin{figure}[t]
\centering
\includegraphics[width=0.95\columnwidth]{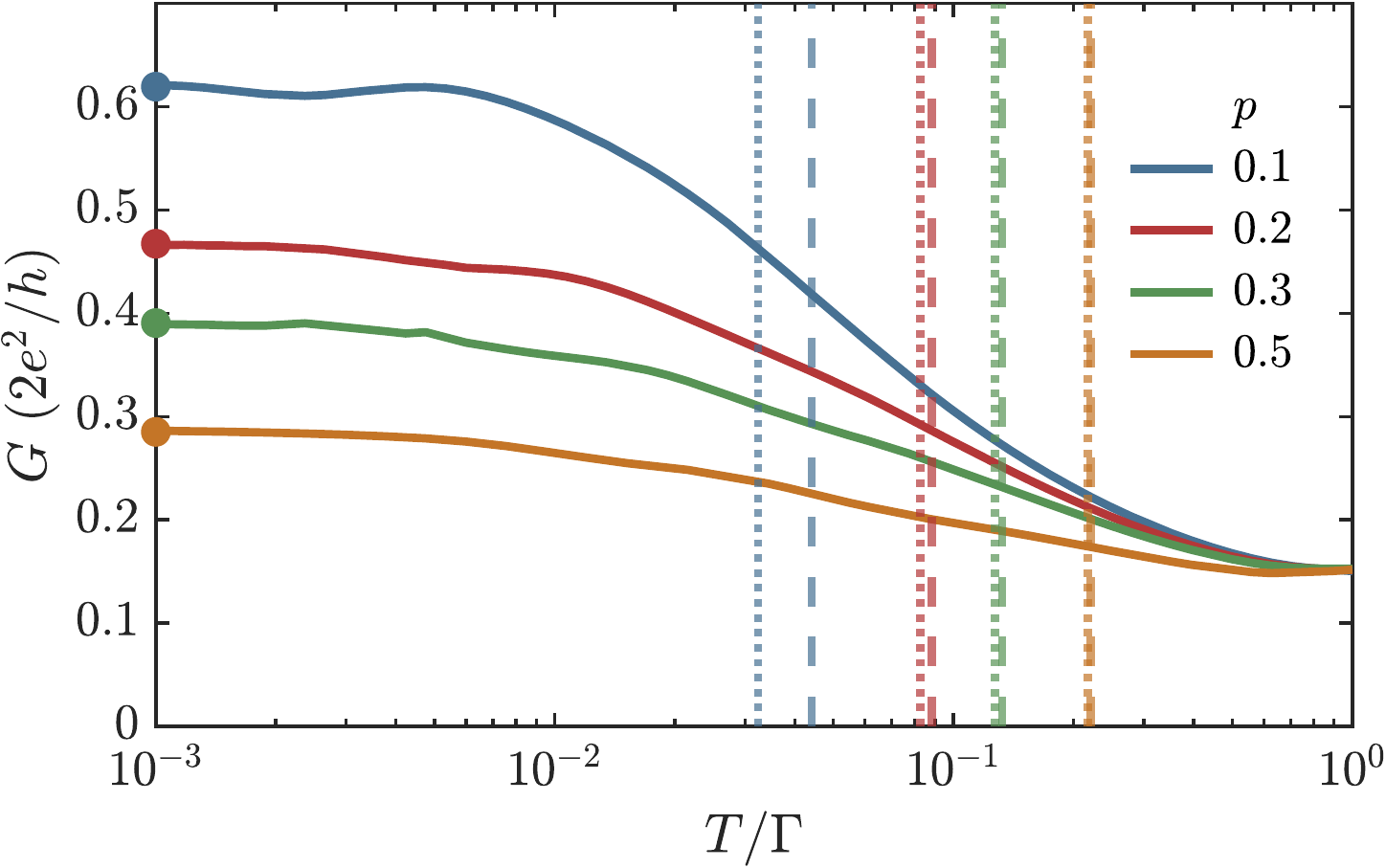}
\caption{
   The temperature dependence of the differential conductance
   $G$ at fixed bias voltage $V=V_{\mathrm{peak}}$
   corresponding to the maximum of the split Kondo peak 
   in \Fig{Fig:detuned}(a) calculated for different spin polarizations,
  using $\e_d=-U/3$ [\Eq{eq:param}],
   where the symbols on the left axis
   replicate the peak in the zero-temperature data in \Fig{Fig:detuned}. 
   The color-matched vertical dotted lines mark the peak bias voltage $V(p)$
   at which conductance is calculated, whereas the dashed
   lines indicate $|\exchp|$. These are roughly located
   where the peak conductance is reduced by about half
   relative to a background conductance due to the
   hybridization side-peaks at energy $\varepsilon_d$.
}
\label{Fig:GvT}
\end{figure}

\subsection{Temperature dependence of split Kondo peak}

In this section we analyze the effect of finite temperature on
the split Kondo resonance. Figure~\ref{Fig:temperature} shows the bias voltage dependence
of the differential conductance for various temperatures
calculated for $\e_d=-U/3$ and $p=0.2$.
One can see that increasing $T$ results in the suppression of
the split Kondo peak, which completely disappears once
the thermal energy exceeds the induced exchange splitting.
Increasing temperature still further overall suppresses the differential conductance.
The suppression of the split-Kondo peak is accompanied
with a weak increase of the conductance at zero bias for
temperatures corresponding to the splitting of the lDOS due to the exchange field, as seen \new{in} the
inset of Fig.~\ref{Fig:temperature}(b).
This can be used to estimate the temperature where the
splitting in the differential conductance disappears. The split-Kondo peak can survive up to a maximum temperature $T_{\mathrm{max}}$ defined as the temperature in which $G(V\rightarrow 0, T) = G(V=V_{\mathrm{peak}}, T)$. For the spin polarization $p=0.2$, we estimate  $T_\mathrm{max} = 2.06 \cdot 10^{-2} \, \Gamma = T_{\mathrm{K},0.2}$.

The $V\rightarrow0$ differential conductance
is equivalent to linear-response in thermal equilibrium.
The latter is readily obtained by NRG, with a direct
comparison shown in the inset of \Fig{Fig:temperature}(b).
Overall, we observe good quantitative agreement.
The points corresponding to the temperatures plotted
in the main panels are marked by the same color-matched symbols (squares).
Since linear response can be efficiently obtained by NRG,
this permits a more dense set of data points in the inset.

The weak increase in the linear-response
conductance for a finite temperature can be explained 
by examining the energy dependence of the 
equilibrium local density of states, i.e., the impurity
spectral function, assuming that this lDOS changes only weakly
at low temperatures $T\lesssim \mathrm{max}(T_{\mathrm{K}},|\exch|)$.
The linear-response conductance $G = \Sigma_\s G_\s$ is 
obtained from the spectral function using $G_\s(T) = \frac{\pi e^2}{h}
\int_{-D}^{D} d\w\, \Gamma_\s A_\s(\w) [-f'(\omega)]$
\cite{Meir1991Jun},
where $A_\s(\w) = -\frac{1}{\pi}\mathrm{Im}\,G_\s(\w) $
is the spin-resolved spectral function based on 
the retarded impurity Green's function $G_\s(\w)$,
and $f'(\w)$ is the derivative of the Fermi function at
temperature $T$. 
Now if the exchange field due to polarization
is sufficiently strong, $|\exch| \gtrsim T_{\mathrm{K}}$,
this will already split the spin-averaged lDOS at equilibrium, 
as shown for $p=0.2$ in \Fig{Fig:Spec}. When temperature is increased, the transport
window widens, and thus encompasses more weight 
from the split peaks. Assuming that the lDOS
only changes weakly by turning on a small temperature
$T\lesssim \mathrm{max}(T_{\mathrm{K}},|\exch|)$,
the contributions from the peak in the spectral function 
around $\omega\approx |\exch|$ 
will therefore increase the linear-response conductance
up to $T\lesssim |\exch|$, where it reaches
a maximum before it starts to decrease.

An explicit temperature dependence of the split-Kondo
peak conductance for a few selected values of spin polarization is shown in
\Fig{Fig:GvT}. This figure is determined at finite bias
voltage $V(p)\approx|\exchp|$, i.e. at the voltage corresponding to
the location of the split Kondo peak $V_{\mathrm{peak}}$ shown in \Fig{Fig:detuned}.
As seen by the vertical markers in \Fig{Fig:GvT},
$V$ agrees well with $|\exch|$ for large polarization $p$,
but clearly starts to differ for smaller $p$, given that
there is no peak at finite $V$ for $p\lesssim 0.0834$.
By starting from the peak conductance,
one can now clearly see in \Fig{Fig:GvT} the decrease of the
remaining side Kondo resonance as the temperature
increases. The logarithmic decrease in the split-Kondo peak conductance at higher temperatures has been experimentally observed in the Fig. 3a of Ref.\cite{Pasupathy2004Oct}.
In the case of $p=0.1$, the split Kondo peak
just emerged, having
$\exch^{p=0.1} \approx T_{\mathrm{K}}$, 
as can be observed from Fig.~\ref{Fig:detuned} and the
vertical blue lines in Fig.~\ref{Fig:GvT}. 
Hence, we can see a slight non-monotonic behavior arising from the interplay
between the Kondo effect and the exchange field. 
More generally, one can infer from Figs.~\ref{Fig:temperature} and \ref{Fig:GvT}, 
for the split-Kondo regime, i.e., sufficiently strong polarization $p$ with $|\exch| >T_{\mathrm{K}}$,
that $G_V$ vs. $T$ [$G_T$ vs. $V$] 
will exhibit a non-monotonic behavior if $T < |\exch|$
[$V < |\exch|$], yet a monotonic  
decay if $T \gtrsim |\exch|$ [or $V \gtrsim |\exch|$], respectively.
We also note that the temperature dependence of the
non-equilibrium differential conductance at $V\approx |\exch|$
does not show a universal dependence. This can be understood
by realizing that the system is then out of the Kondo regime.

\section{Summary}
\label{Sec:Summary}

In this paper we have studied the non-equilibrium spin-resolved transport through a quantum dot coupled to ferromagnetic leads, while treating the correlations exactly. When the dot level is at the particle-hole symmetry point, we have shown that the Kondo resonance can be observed for any value of spin polarization $p$, but the Kondo energy scale in the bias potential $V_{\mathrm{K}}$ reduces with increasing spin polarization. However, when the dot level is detuned out of the particle-hole symmetry point, we have observed the emergence of an exchange field $\exch$ in the system, which splits the zero-bias conductance peak when it is comparable or larger than the Kondo energy scale. A finite value of magnetic field $B \approx \exch$ was able to restore the Kondo resonance in such system. Moreover, we have determined the temperature dependence of the split Kondo peak and showed that the character of this dependence depends on the ratio of exchange field to the Kondo energy scale. Our work provides benchmark results for the nonequilibrium spintronic transport through quantum impurity systems in the presence of ferromagnetic leads.


\begin{acknowledgments}
    This work was supported by the Polish National Science
    Centre from funds awarded through the decision Nos.~2017/27/B/ST3/00621 and 2021/41/N/ST3/02098.
    We would also like to acknowledge the support by the project ``Initiative of Excellence - Research University'' from funds awarded through Decision no: 003/13/UAM/0016. AW was supported by the
    U.S. Department of Energy (DOE)
    Office of Basic Energy Sciences (BES),
    Materials Sciences and Engineering Devision.
\end{acknowledgments}

\appendix
\section{The hybrid NRG-tDMRG thermofield quench approach}
\label{app:Appendix}

This appendix provides more details on the hybrid NRG-tDMRG thermofield quench method \cite{Schwarz2018Sep} used to calculate the spin-resolved transport properties of the system in non-equilibrium settings.

\subsection{Thermofield treatment of the leads}

To describe the leads we use the thermofield approach
\cite{Barnett1987Feb,Das2000Apr,deVega2015Nov},
in which an auxiliary Hilbert space, equivalent to the lead Hilbert space, but decoupled from the system, is introduced to the lead Hamiltonian, effectively doubling the Hilbert space.
This allows us to simplify the computational problem,
since the decoupled modes of thermal leads can be expressed
as simple product states.  More importantly, thermofield approach enables the description of the thermal states as pure states,
which can be then time-evolved
within the matrix product state framework.

A pure state $\ket{\Omega}$ is defined on this enlarged space such that the thermal expectation value of an observable $A$ in the original physical Hilbert space can be obtained from the enlarged space using
$\langle A \rangle = \langle\Omega|A|\Omega\rangle$, where
the state $\ket{\Omega}$ is defined as
\be
\ket{\Omega} = \prod_q (\sqrt{1-f_q}\ket{0,1}_q + \sqrt{f_q}\ket{1,0}_q).
\ee
Here, the composite index corresponds to $q\equiv \{\alpha,k,\s\}$,
$f_q \equiv f_\alpha (\e_{\alpha k \sigma})$,
and the Fock states, $\ket{0,1}_q$
and $\ket{1,0}_q$, which act as the basis for the new Hilbert space,
are defined as, $c_{q1}\ket{0,1}_q=\ck_{q2}\ket{0,1}_q=\ck_{q1}\ket{1,0}_q=c_{q2}\ket{1,0}_q=0$.
We define the modes $\tilde{c}_{qj}$ in a rotated basis such that, $\ket{\tilde{0},\tilde{1}}_q=\sqrt{1-f_q}\ket{0,1}_q + \sqrt{f_q}\ket{1,0}_q$,
using the transformation,
\be
\begin{pmatrix}
\tilde{c}_{q1}\\
\tilde{c}_{q2}
\end{pmatrix} 
=
\begin{pmatrix}
\sqrt{1-f_q} & \sqrt{f_q}\\
\sqrt{f_q} & \sqrt{1-f_q}
\end{pmatrix}
\begin{pmatrix}
c_{q1}\\
c_{q2}
\end{pmatrix}.
\ee
With this transformation, the initial pure product state $\ket{\Omega}$
is such that $\tilde{c}_{q1}\ket{0,1}_q=\tilde{c}^\dag_{q2}\ket{0,1}_q=0$,
which essentially results in one set of modes ($j=2$) to be fully occupied,
while the rest ($j=1$) is empty. The fully filled (empty) states in the new basis resemble the particle (hole) description of the lead Hamiltonian.
\new{The particles and holes will be recombined later for the NRG part of the calculations
	but treated separately for the tDMRG time evolution as described later.}
\subsection{The hybrid NRG-tDMRG time evolution
\label{sec:NRG:tDMRG}}

The hybrid NRG-tDMRG approach we employ combines
the strong assets of both NRG and DMRG, namely,
the ability of NRG to resolve logarithmic energy scales
and the ability of DMRG to describe nonequilibrium situations
at energy scales close to the bandwidth. One fundamental difference between the both methods is that while NRG is fundamentally based
on logarithmic discretization, DMRG studies have found
incredible success based on a linear discretization of the lead energy continuum. The energy scales that distinguish the regimes of implementation of these methods are denoted by the transport window (TW), which is determined by 
a difference in the electrochemical potentials of the leads,
$f_L(\w) \ne f_R(\w)$.
Assuming that the lead levels far from the TW are essentially in equilibrium,
we implement a logarithmic discretization scheme outside the transport window
in order to later treat them with the aid of the NRG.
On the other hand, the energies inside the TW are discretized linearly to be compatible with the DMRG formalism.
The discretized energy intervals are denoted by $E_k$ and defined as,
\begin{equation*}
E_k(x)= 
\begin{cases}
\delta \cdot x, & \text{for} \mid x \mid \leq D^*/\delta\\
\frac{\delta \cdot \text{sinh}(\text{ln}(\Lambda)(x \mp \frac{D^*}{\delta}))}{\text{log}(\Lambda)} \mp \delta\cdot D^*, & \text{for} \ x \lessgtr \pm D^*/\delta,
\end{cases}
\end{equation*}
where $\delta$ and $\Lambda$ are the linear and logarithmic discretization parameters, respectively.
The energy levels outside the TW are treated using the numerical renormalization group method,
giving rise to a renormalized impurity (RI) with a reduced effective bandwidth $2D^*$.
As a result of the thermofield transformation in the linear sector,
the system can be effectively described as renormalized
impurity coupled to two chains,
corresponding to the tridiagonalized chains of the particle and hole modes.

The Hamiltonians, $H_\text{lead}$ and $H_\text{hyb}$, transform according to the aforementioned rotation as,
\bea
\mathcal{H}_\text{lead} = H_\text{lead} + H_\text{aux} &=& \sum_{qj}\e_q\ck_{qj}c_{qj}=\sum_{qj}\e_q\Tilde{c}^\dag_{qj}\Tilde{c}_{qj},\nonumber\\
H_\text{hyb}&=&\sum_{qj} (\Tilde{v}_{qj}\dk_{\s}\Tilde{c}_{qj} + \text{H.c.}),
\eea
where $j \in \{1,2\}$ and the transformed couplings
${\tilde{v}_{q1} = v_q\sqrt{1-f_q}}$ and ${\tilde{v}_{q2}=v_q\sqrt{f_q}}$.
\new{After the transformation, we recombine the particles and holes
	in the logarithmically discretized regime through another tridiagonalization in order to apply NRG.}
Furthermore, we recombine the transformed left and right lead modes
so that one set of modes decouples from the system,
as is common in the case of equilibrium NRG studies \cite{Bulla2008Apr}.

We perform a second order Trotter time evolution on the initial state of the system,
$\ket{\psi_{\text{ini}}} = \ket{\phi_{\text{ini}}} \otimes \ket{\Omega}$,
during which the coupling between the linear and logarithmic sectors is switched on over a finite time interval.
Here, $\ket{\phi_{\text{ini}}}$ is the initial state of the RI
and $\ket{\Omega}$ is the pure product state of the linear sector.
We calculate the symmetrized current
\begin{eqnarray}
   J = J_L - J_R
\label{eq:J:sym}
\end{eqnarray}
at each time step of the system's evolution,
where $J_L$ ($J_R$) is defined as the current flowing
from the left (right)
lead to the impurity and $J_\alpha=\sum_\s J_{\alpha \s}$.
The system is time-evolved until
the relevant observables start to fluctuate around a mean value and a nonequilibrium steady state is reached.
We evaluate our main quantity of
interest---the current---as the mean of the symmetrized current
over a finite time interval where the system shows steady state behavior. The averaging time window is chosen by scanning 
 through the current dynamics to find the one
 with least error around the mean value.
The corresponding differential conductance $G=d J(V)/d V$
is calculated from the mean symmetrized current.
Both NRG and tDMRG calculations are implemented
in the matrix product state framework
\cite{Andreas2012}.
In calculations we assume $\Lambda = 2.5$ and  $\delta = 0.0625 D^*$.

\bibliography{noneq-spinvalve}

\end{document}